\documentclass[aps,pra,showpacs,superscriptaddress,preprint]{revtex4}
\usepackage{amsmath}
\usepackage{graphicx}

\catcode`ð=\active
 \defð{\u{g}}
 \catcode`Ð=\active
 \defÐ{\u{G}}
 \catcode`Ý=\active
\defÝ{\. I}
 \catcode`ö=\active
\defö{\"{o}}
 \catcode`Ö=\active
 \defÖ{\"O}
 \catcode`ü=\active
 \defü{\"{u}}
 \catcode`Ü=\active
 \defÜ{\"{U}}
 \catcode`Þ=\active
\defÞ{\c{S}}
 \catcode`þ=\active
 \defþ{\c{s}}
 \catcode`ý=\active
 \defý{{\i}}
 \catcode`ç=\active
\defç{\d{c}}
 \catcode`Ç=\active
\defÇ{\d{C}}

\begin{document}

\title{Any $l$-state solutions of the Woods-Saxon potential in arbitrary
dimensions within the new improved quantization rule }
\author{Sameer M. Ikhdair}
\email[E-mail: ]{sikhdair@neu.edu.tr}
\affiliation{Physics Department, Near East University, Nicosia, North Cyprus, Turkey}
\author{Ramazan Sever}
\email[E-mail: ]{sever@metu.edu.tr}
\affiliation{Physics Department, Middle East Technical University, 06800, Ankara, Turkey}
\date{%
\today%
}

\begin{abstract}
The approximated energy eigenvalues and the corresponding eigenfunctions of
the spherical Woods-Saxon effective potential in $D$ dimensions are obtained
within the new improved quantization rule for all $l$-states. The Pekeris
approximation is used to deal with the centrifugal term in the effective
Woods-Saxon potential. The inter-dimensional degeneracies for various
orbital quantum number $l$ and dimensional space $D$ are studied. The
solutions for the Hulth\'{e}n potential, the three-dimensional ($D=3$), the $%
s$-wave ($l=0$) and the cases are briefly discussed.

\end{abstract}

\maketitle

\newpage

\section{Introduction}

In recent years, the analytic bound state solution of the hyperradial Schr%
\"{o}dinger equation, in any arbitrary spatial dimension $(D\geq 2),$ with
several potential models has received increasing attention in the literature
[1]. There are only a few potentials for which the radial Schr\"{o}dinger
equation can be solved explicitly for all $n$ and $l$ quantum numbers$.$ One
of these exactly solvable potentials is the Woods-Saxon (WS) potential [2]
which has been firstly solved within the Schr\"{o}dinger equation for $s$%
-states ($l=0)$ by Bose [3,4]. However, the three-dimensional radial Schr%
\"{o}dinger equation for the spherical WS potential cannot be solved
analytically for $l\neq 0$ states because of the centrifugal term $\sim
r^{-2}.$ This potential was used as a major part of nuclear shell model to
obtain the nuclear energy level spacing and properties of electron
distributions in atoms, nuclei and atomic clusters [5-7]. It was used to
describe the interaction of a nucleon (neutron) with a heavy nucleus [8] and
also for optical potential model in elastic scattering of some ions with
heavy target in low range of energies [9].

So far, numerous attempts have been developed to calculate the bound-state
energies of the WS potential in the framework of the Schr\"{o}dinger
equation by means of various methods; the Nikiforov-Uvarov method [10-12]
and the shape invariance and Hamiltonian hierarchy [13]. However, the
solutions in [10,11] are at most valid for $R_{0}=0,$ in which the WS
potential turns out to become the Rosen-Morse-type potential. The
two-component approach to the one-dimensional Dirac equation with WS
potential is applied to obtain the scattering and bound-state solutions
[14]. The exact solution of the relativistic Dirac equation was obtained for
a single particle with spin and pseudospin symmetry moving in a central WS
potential using two-component spinors [15-17]. An approach has been
developed to obtain the bound state solutions of the scattering of a
Klein-Gordon particle by a WS potential [18]. Recently, the approximated
eigenenergies and eigenfunctions of the Dirac equation for the Woods-Saxon
potential and a tensor potential with the arbitrary spin-orbit coupling
quantum number $\kappa $ under pseudospin and spin symmetry have been
obtained [19].

Recently, Ma and Xu have proposed an improved quantization rule (IQR) and
shown its power in calculating the energy levels of all bound states for
some solvable quantum systems [20,21]. The method has been shown to be
effective for calculating the bound state solutions of the Schr\"{o}dinger
and Dirac wave equations with a spherically symmetric potential [1,22-26].
So far, it has been applied, with great success, to study a great number of
potentials like the rotating Morse [22,23], the Kratzer-type [24], the
trigonometric Rosen-Morse [25], the hyperbolic and the second P\"{o}%
schl-Teller-like potentials [26] and the Hulth\'{e}n potential [1]
and so forth. Very recently, Gu and Sun [1] have extended the
application of the IQR to the solution of the $D$-dimensional
Schr\"{o}dinger equation with the Hulth\'{e}n potential for $l\neq
0$ using the usual approximation to deal with the centrifugal term
[27-29].

In this paper, we present a new systematical approach to solve the Schr\"{o}%
dinger equation in $D$-dimensions with WS potential for it's energy
eigenvalues and the corresponding eigenfunctions by means of the IQR method
using the Pekeris approximation scheme [30-32] to deal with the centrifugal
term. This alternative approach has recently shown its accuracy in
calculating the analytic spectrum of the Hulth\'{e}n potential for $l\neq 0$
[1]. Further, we give analytic tests using energy calculations for
interdimensional degeneracy, i.e., $(n,l,D)\rightarrow (n,l\pm 1,D\mp 2)$
corresponding to the confined $D=2-4$ dimensional Woods-Saxon potential.

This paper is organized as follows. In Sec. 2, the IQR method is reviewed
and extended to any arbitrary dimension $(D\geq 2)$. In Sec. 3, the $D$%
-dimensional ($D\geq 2$) Schr\"{o}dinger equation is solved by this method
with arbitrary $l$-states to obtain all the energy states of the Woods-Saxon
potential using the Pekeris approximation. In Sec. 4, we calculate the
corresponding hyperradial wave functions of the Woods-Saxon potential. In
Sec. 5, the interdimensional degeneracy is introduced. Finally, some
concluding remarks are given in Sec. 6.

\section{Improved quantization rule}

A brief outline to the improved quantization rule is presented with an
extension to the $D$-dimensional space ($D\geq 2$). The details can be found
in Refs. [20,21]. The IQR has recently been proposed to solve exactly the
one-dimensional ($1D$) Schr\"{o}dinger equation:

\begin{equation}
\psi {}^{\prime \prime }(x)+k(x)^{2}\psi (x)=0,\text{ \ }k(x)=\frac{\sqrt{%
2\mu \left[ E-V(x)\right] }}{\hbar },
\end{equation}%
where the prime denotes the derivative with respect to the variable $x.$
Here $\mu $ is the reduced mass of the two interacting particles, $k(x)$ is
the momentum and $V(x)$ is a piecewise continuous real potential function of
$x.$ The Schr\"{o}dinger equation is equivalent to the Riccati equation%
\begin{equation}
\phi {}^{\prime }(x)+\phi (x)^{2}+k(x)^{2}=0,
\end{equation}%
where $\phi (x)=\psi {}^{\prime }(x)/\psi (x)$ is the logarithmic derivative
of wave function $\psi (x).$ Due to the Sturm-Liouville theorem, the $\phi
(x)$ decreases monotonically with respect to $x$ between two turning points,
where $E\geq V(x).$ Specifically, as $x$ increases across a node of the wave
function $\psi (x),$ $\phi (x)$ decreases to $-\infty ,$ jumps to $+\infty ,$
and then decreases again.

Moreover, Ma and Xu [20,21] have generalized this exact quantization rule to
the three-dimensional $\left( 3D\right) $ radial Schr\"{o}dinger equation
with spherically symmetric potential by simply making the replacements $%
x\rightarrow $ $r$ and $V(x)\rightarrow V_{\text{eff}}(r)$:

\begin{equation}
\int\limits_{r_{A}}^{r_{B}}k(r)dr=N\pi
+\int\limits_{r_{A}}^{r_{B}}k{}^{\prime }(r)\frac{\phi (r)}{\phi {}^{\prime
}(r)}dr,\text{ }k(r)=\frac{\sqrt{2\mu \left[ E_{n,l}-V_{\text{eff}}(r)\right]
}}{\hbar },
\end{equation}%
where $r_{A}$ and $r_{B}$ are two turning points determined from the
relation $E_{n,l}=V_{\text{eff}}(r),$ $N=n+1$ is the number of nodes of \ $%
\phi (r)$ in the region $E_{n,l}\geq V_{\text{eff}}(r)$ and it is larger by
one than the number of nodes of wave function $\psi (r).$ The first term $%
N\pi $ is the contribution from the nodes of the logarithmic derivative of
wave function, and the second term in (3) is called the quantum correction.
It is found that, for all well-known exactly solvable quantum systems, this
quantum correction is independent of the number of nodes of wave function of
the system. This means that it is enough to consider the ground state in
calculating the quantum correction, i.e.,
\begin{equation}
Q_{c}=\int\limits_{r_{A}}^{r_{B}}k_{0}{}^{\prime }(r)\frac{\phi _{0}(r)}{%
\phi _{0}{}^{\prime }(r)}dr=\pi \nu .
\end{equation}%
The quantization rule still holds for Schr\"{o}dinger equation with
spherically symmetric potential in $D$ dimensions. In what follows, we shall
employ this method to solve the Schr\"{o}dinger equation in $D$-dimensions
with WS potential using the Pekeris approximation to deal with the
centrifugal term.

\section{Eigenvalues of the Woods-Saxon potential}

The $D$ dimensional Schr\"{o}dinger equation with spherically symmetric
potential $V(r)$ for arbitrary $l$-state takes the form
\begin{equation}
\left( -\frac{\hbar ^{2}}{2\mu }\nabla _{D}^{2}+V(r)-E_{n,l}\right) \psi
_{n,l,m}(r,\Omega _{D})=0,\text{ }
\end{equation}%
where the representation of the Laplacian operator $\nabla _{D}^{2},$ in
spherical coordinates, is
\begin{equation}
\nabla _{D}^{2}=\frac{\partial ^{2}}{\partial r^{2}}+\frac{\left( D-1\right)
}{r}\frac{\partial }{\partial r}-\frac{l\left( l+D-2\right) }{r^{2}},
\end{equation}%
and%
\begin{equation}
\psi _{n,l,m}(r,\Omega _{D})=\psi _{n,l}(r)Y_{l}^{m}(\Omega _{D}),\text{ }%
\psi _{n,l}(r)=r^{-(D-1)/2}u(r),
\end{equation}%
where $Y_{l}^{m}(\Omega _{D})$ is the hyperspherical harmonics. The wave
functions $\psi _{n,l,m}(r,\Omega _{D})$ belong to the energy eigenvalues $%
E_{n,l}$ and $V(r)$ is taken, in the present study, as the
Woods-Saxon potential in the configuration space and $r$ represents
the $D$-dimensional intermolecular distance $\left(
\sum\limits_{i=1}^{D}x_{i}^{2}\right) ^{1/2}.$

Further, substituting Eqs. (6) and (7) into Eq. (5) yield the wave equation
satisfying the radial wave function $u(r)$ in a simple analogy to the $2D$
and $3D$ radial Schr\"{o}dinger equation
\begin{equation}
\frac{d^{2}u(r)}{dr^{2}}+\frac{2\mu }{\hbar ^{2}}\left[ E_{n,l}-V_{eff}(r)%
\right] u(r)=0,
\end{equation}%
where $E_{n,l}$ is the bound state energy of the system and $V_{eff}(r)$ is
the deformed Woods-Saxon [2] effective potential in $D$ dimensions defined
by
\begin{equation}
V_{eff}(r)=-\frac{V_{0}e^{-\frac{\left( r-R_{0}\right) }{a}}}{1+qe^{-\frac{%
\left( r-R_{0}\right) }{a}}}+\frac{\lambda }{r^{2}},
\end{equation}%
with the parameter
\begin{equation}
\lambda =\frac{\left( \Lambda ^{2}-1\right) \hbar ^{2}}{8\mu },\text{ }%
\Lambda =2l+D-2.
\end{equation}%
where $R_{0}=r_{0}A^{1/3}$ is the nuclear radius with $r_{0}=1.25$ $fm$ (may
vary by as much as $0.2$ $fm$ depending on the specific nuclide) and $A$ the
atomic mass number of protons $(Z)$ plus number of neutrons $(N)$ in target
nucleus, $V_{0}\approx 50$ $MeV$ (having. dimension of energy) is the
potential well depth and $a\approx 0.5-0.6$ $fm$ is the length representing
the surface thickness that usually adjusted to the experimental values of
ionization energies [12]. In the case of negative eigenenergies (i.e., when $%
E_{n,l}\in \lbrack -50$ $MeV,0])$ we have the well-known bound-states
problem while in the case of positive eigenenergies (i.e., when $E_{n,l}\in
\lbrack 0,1000$ $MeV])$ we have the well-known resonance problem. It should
be noted that the deformation parameter $q$ is a real parameter and can be
taken equal to $1$ and $-1$ for the Woods-Saxon and the Hulth\'{e}n
potentials, respectively in the calculations$.$ The radial wave function $%
u(r)$ satisfying Eq. (8) should be normalizable and finite near $r=0$ and $%
r\rightarrow \infty $ for the bound-state solutions. The wave equation (8)
with the Woods-Saxon potential is an exactly solvable problem for $l=0$ ($s$%
-wave) [3,4,15], however, it cannot be solved analytically for $\lambda \neq
0$ because of the orbital coupling term $\lambda r^{-2}$. Therefore, to
solve Eq. (8) analytically, we must use the Pekeris approximation [30-32] to
deal with the centrifugal term. By inserting the conversions $x=\left(
r-R_{0}\right) /R_{0}$ and $\alpha =R_{0}/a,$ where $r\in (0,\infty
)\rightarrow x\in (-1,\infty ),$ the following exponential form can be used
instead of the centrifugal term%
\begin{equation}
\frac{1}{r^{2}}\approx \frac{1}{R_{0}^{2}}\left[ d_{0}+d_{1}\frac{e^{-\alpha
x}}{1+qe^{-\alpha x}}+d_{2}\frac{e^{-2\alpha x}}{\left( 1+qe^{-\alpha
x}\right) ^{2}}\right] ,
\end{equation}%
which can be expanded around the minimum point $r\approx R_{0}$ (or $x=0$)
only up to the second order as%
\begin{equation}
\frac{1}{r^{2}}\approx \frac{1}{R_{0}^{2}}\left[ d_{0}+\frac{d_{1}}{2}+\frac{%
d_{2}}{4}-\frac{\alpha }{4}\left( d_{1}+d_{2}\right) x+\frac{\alpha ^{2}}{16}%
d_{2}x^{2}-\cdots \right] .
\end{equation}%
where $d_{0},$ $d_{1}$ and $d_{2}$ are coupling constant parameters. It is
worth to note that the above expansion is valid for low rotational energy
states and $q=1.$ The expression $r^{-2}$ can be also expanded around $x=0$
up to the second order term as
\begin{equation}
\frac{1}{r^{2}}=\frac{1}{R_{0}^{2}}\left( 1-2x+3x^{2}-\cdots \right) .
\end{equation}%
Comparing Eqs. (12) and (13), we obtain the constant parameters for the
Woods-Saxon potential as follows%
\begin{equation}
d_{0}=1-\frac{4}{\alpha }+\frac{12}{\alpha ^{2}},\text{ }d_{1}=\frac{8}{%
\alpha }-\frac{48}{\alpha ^{2}},\text{ }d_{2}=\frac{48}{\alpha ^{2}},
\end{equation}%
However, these constant parameters for the Hulth\'{e}n potential take values
as follows%
\begin{equation}
d_{0}=\frac{1}{12},\text{ }d_{1}=1,\text{ }d_{2}=1.
\end{equation}%
Further, by defining%
\begin{equation}
b=\left( \frac{Q\delta }{R_{0}}\right) ^{2},\text{ }\delta ^{2}=\frac{\hbar
^{2}}{2\mu }\left( l+\frac{D-1}{2}\right) \left( l+\frac{D-3}{2}\right) ,
\end{equation}%
where the scaling parameter $Q$ is introduced with the purpose of obtaining
solution for the Hulth\'{e}n potential as well. So we take $Q=1$ for the
Woods-Saxon potential but $Q=\alpha $ and $R_{0}=1$ for the Hulth\'{e}n
potential.

Equation (8) can be rewritten as%
\begin{equation}
\frac{d^{2}u(x)}{dx^{2}}+\frac{2\mu R_{0}^{2}}{\hbar ^{2}}\left[
E_{n,l}+V_{0}\frac{e^{-\alpha x}}{1+qe^{-\alpha x}}-b\left( d_{0}+d_{1}\frac{%
e^{-\alpha x}}{1+qe^{-\alpha x}}+d_{2}\frac{e^{-2\alpha x}}{\left(
1+qe^{-\alpha x}\right) ^{2}}\right) \right] u(x)=0,\text{ }
\end{equation}%
where $n$ and $l$ signify the radial and orbital angular quantum numbers,
respectively.

We now study this quantum system through the improved exact quantization
rule. At first, we introduce a new variable%
\begin{equation}
y=\frac{e^{-\alpha x}}{1+qe^{-\alpha x}},\text{ }\frac{dy}{dx}=-\alpha
y(1-qy),\text{ }\frac{dy}{dr}=-\frac{\alpha }{R_{0}}y(1-qy),
\end{equation}%
where $y\in (0,\frac{e^{\alpha }}{1+qe^{\alpha }}).$ Overmore, the turning
points $y_{A\text{ }}$ and $y_{B\text{ }}$ are determined by solving $%
V_{eff}(y)=bd_{2}y^{2}+\left( bd_{1}-V_{0}\right) y+bd_{0}=E_{n,l}$ as
follows:
\begin{subequations}
\begin{equation}
y_{A\text{ }}=\frac{V_{0}}{2bd_{2}}-\frac{d_{1}}{2d_{2}}-\frac{1}{2bd_{2}}%
\sqrt{\left( V_{0}-bd_{1}\right) ^{2}+4bd_{2}\left( E_{nl}-bd_{0}\right) },
\end{equation}%
\begin{equation}
y_{B\text{ }}=\frac{V_{0}}{2bd_{2}}-\frac{d_{1}}{2d_{2}}+\frac{1}{2bd_{2}}%
\sqrt{\left( V_{0}-bd_{1}\right) ^{2}+4bd_{2}\left( E_{nl}-bd_{0}\right) },
\end{equation}%
with the properties
\end{subequations}
\begin{equation}
y_{A\text{ }}+y_{B\text{ }}=\frac{V_{0}}{bd_{2}}-\frac{d_{1}}{d_{2}},\text{ }%
y_{A\text{ }}y_{B\text{ }}=-\frac{1}{bd_{2}}\left( E_{nl}-bd_{0}\right) .
\end{equation}%
The momentum $k(y)$ between two turning points is expressed as
\begin{equation*}
k(y)=\frac{\sqrt{2\mu }}{\hbar }\sqrt{bd_{2}}\sqrt{-y^{2}+\left( \frac{V_{0}%
}{bd_{2}}-\frac{d_{1}}{d_{2}}\right) y+\frac{1}{bd_{2}}\left(
bd_{0}-E_{nl}\right) }
\end{equation*}%
\begin{equation}
=\frac{\sqrt{2\mu }}{\hbar }\sqrt{bd_{2}}\sqrt{\left( y_{B\text{ }}-y\right)
\left( y-y_{A\text{ }}\right) },
\end{equation}%
\begin{equation}
\frac{dk(y)}{dy}=\frac{\sqrt{2\mu }}{2\hbar }\sqrt{bd_{2}}\left( \sqrt{\frac{%
y_{B\text{ }}-y}{y-y_{A\text{ }}}}-\sqrt{\frac{y-y_{A\text{ }}}{y_{B\text{ }%
}-y}}\right) ,
\end{equation}%
The Riccati equation (2) now becomes
\begin{equation}
-\frac{\alpha }{R_{0}}y(1-qy)\frac{d\phi _{0}(y)}{dy}=-\frac{2\mu }{\hbar
^{2}}\left[ E_{0}-bd_{2}y^{2}+\left( V_{0}-bd_{1}\right) y-bd_{0}\right]
-\phi _{0}(y)^{2},
\end{equation}%
having the only possible first order polynomial solution satisfying%
\begin{equation}
\phi _{0}(r)=c_{1}y+c_{2},\text{ }\frac{\text{d}\phi _{0}{}(r)}{dr}=-\frac{%
\alpha }{R_{0}}c_{1}y(1-qy),\text{ }c_{1}>0.
\end{equation}%
where we have used $\phi _{0}(r)\equiv \phi _{0}(y).$ Substituting $\phi
_{0}(y)$ into Eq. (23), one has the ground state wave function and energy
eigenvalue solutions%
\begin{equation}
\left\{
\begin{array}{c}
\phi _{0}(y)=\frac{\alpha }{R_{0}}my+\left( \frac{\alpha }{2R_{0}}-\frac{%
2\mu }{\hbar ^{2}}\frac{\left( V_{0}-bd_{1}\right) }{2m\alpha }R_{0}\right) ,
\\
m=-\frac{q}{2}\left( 1\pm \sqrt{1+\frac{8\mu }{\hbar ^{2}}\frac{Q^{2}\delta
^{2}d_{2}}{\alpha ^{2}q^{2}}}\right) ,\text{ }m\geq 0, \\
\widetilde{E}_{n=0}=E_{0}-bd_{0}=-\frac{\hbar ^{2}}{2\mu }\left[ \frac{%
\alpha }{2R_{0}}-\frac{2\mu }{\hbar ^{2}}\frac{\left( V_{0}-bd_{1}\right) }{%
2m\alpha }R_{0}\right] ^{2}.%
\end{array}%
\right.
\end{equation}%
The new quantum number is chosen as $m=-\frac{q}{2}\left( 1-\sqrt{1+\frac{%
8\mu }{\hbar ^{2}}\frac{Q^{2}\delta ^{2}d_{2}}{\alpha ^{2}q^{2}}}\right) $
for the Woods-Saxon potential and $m=-\frac{q}{2}\left( 1+\sqrt{1+\frac{8\mu
}{\hbar ^{2}}\frac{Q^{2}\delta ^{2}d_{2}}{\alpha ^{2}q^{2}}}\right) $ for
the Hulth\'{e}n potential. After a lengthy algebra but straightforward, we
can calculate the integral of the quantum correction (4) based on the ground
state as%
\begin{equation}
Q_{c}=-\pi \left( \frac{\sqrt{2\mu d_{2}}}{\hbar }\frac{\delta }{q}\frac{Q}{%
\alpha }+\frac{m}{q}+1\right) .
\end{equation}%
The integral of the momentum $k(r)$ in the quantization rule (3) is
calculated as%
\begin{equation*}
\int\limits_{r_{A}}^{r_{B}}k(r)dr=-\frac{\sqrt{2\mu }}{\hbar }\sqrt{\frac{%
bd_{2}}{q^{2}}}\frac{R_{0}}{\alpha }\int\limits_{z_{A}}^{z_{B}}\left( \frac{%
\sqrt{\left( y-y_{A\text{ }}\right) \left( y_{B\text{ }}-y\right) }}{y}+q%
\frac{\sqrt{\left( y-y_{A\text{ }}\right) \left( y_{B\text{ }}-y\right) }}{%
1-qy}\right) dy
\end{equation*}%
\begin{equation}
=\pi \frac{\sqrt{2\mu }}{\hbar }\frac{R_{0}}{\alpha }\left( \sqrt{\frac{%
bd_{2}}{q^{2}}}+\sqrt{bd_{0}-E_{n,l}}-\sqrt{b\left( d_{0}+\frac{d_{1}}{q}+%
\frac{d_{2}}{q^{2}}\right) -\frac{V_{0}}{q}-E_{n,l}}\right) .
\end{equation}%
Using the relations (26) and (27), the improved quantization rule (3) turn
out to be%
\begin{equation*}
\pi \frac{\sqrt{2\mu }}{\hbar }\frac{R_{0}}{\alpha }\left( \sqrt{\frac{bd_{2}%
}{q^{2}}}+\sqrt{bd_{0}-E_{n,l}}-\sqrt{b\left( d_{0}+\frac{d_{1}}{q}+\frac{%
d_{2}}{q^{2}}\right) -\frac{V_{0}}{q}-E_{n,l}}\right)
\end{equation*}%
\begin{equation}
=\pi \left( n-\frac{\sqrt{2\mu d_{2}}}{\hbar }\frac{\delta }{q}\frac{Q}{%
\alpha }-\frac{m}{q}\right) .
\end{equation}%
Thus, based on the improved quantization rule we can finally obtain the
approximated energy levels $E_{nl}$ valid for both the Woods-Saxon and the
Hulth\'{e}n potentials in the $D$-dimensions,%
\begin{equation*}
E_{n,l}^{(D)}=\frac{\delta ^{2}Q^{2}}{R_{0}^{2}}d_{0}-\frac{\hbar ^{2}}{2\mu
}\frac{R_{0}^{2}}{\alpha ^{2}}
\end{equation*}%
\begin{equation}
\times \left[ \frac{\alpha ^{2}\left( 2n+1\mp \sqrt{1+\frac{8\mu }{\hbar ^{2}%
}\frac{Q^{2}\delta ^{2}}{q^{2}\alpha ^{2}}d_{2}}\right) }{4R_{0}^{2}}-\frac{%
\frac{2\mu }{\hbar ^{2}}\left( qd_{1}+d_{2}\right) \frac{\delta ^{2}Q^{2}}{%
q^{2}}-\frac{2\mu }{\hbar ^{2}}\frac{V_{0}}{q}R_{0}^{2}}{R_{0}^{2}\left(
2n+1\mp \sqrt{1+\frac{8\mu }{\hbar ^{2}}\frac{Q^{2}\delta ^{2}}{q^{2}\alpha
^{2}}d_{2}}\right) }\right] ^{2},
\end{equation}%
where $n,l=0,1,2,\cdots .$ Further, by setting $Q=1$ and the deformation
parameter $q=1$ in the above equation$,$ we get the energy spectrum of the
Woods-Saxon potential as
\begin{equation}
E_{n,l}^{(D)}=\frac{\delta ^{2}}{R_{0}^{2}}d_{0}-\frac{\hbar ^{2}a^{2}}{2\mu
}\times \left[ \frac{\left( 2n+1-\sqrt{1+\frac{8\mu }{\hbar ^{2}}\frac{%
\delta ^{2}}{\alpha ^{2}}d_{2}}\right) }{4a^{2}}-\frac{\frac{2\mu }{\hbar
^{2}}\frac{\delta ^{2}}{R_{0}^{2}}\left( d_{1}+d_{2}\right) -\frac{2\mu }{%
\hbar ^{2}}V_{0}}{\left( 2n+1-\sqrt{1+\frac{8\mu }{\hbar ^{2}}\frac{\delta
^{2}}{\alpha ^{2}}d_{2}}\right) }\right] ^{2},
\end{equation}%
In three-dimensions ($D=3$), it can be reduced to the form%
\begin{equation}
E_{n,l}^{(D)}=\frac{\hbar ^{2}l\left( l+1\right) }{2\mu R_{0}^{2}}d_{0}-%
\frac{\hbar ^{2}a^{2}}{2\mu }\left[ \frac{\left( 2n+1-\sqrt{1+4l\left(
l+1\right) \frac{d_{2}}{\alpha ^{2}}}\right) }{4a^{2}}-\frac{\frac{1}{%
R_{0}^{2}}l\left( l+1\right) \left( d_{1}+d_{2}\right) -\frac{2\mu }{\hbar
^{2}}V_{0}}{\left( 2n+1-\sqrt{1+4l\left( l+1\right) \frac{d_{2}}{\alpha ^{2}}%
}\right) }\right] ^{2}.
\end{equation}%
which is essentially same as that obtained by other Nikiforov-Uvarov method
[33].

On the other hand, the case when we set the parameters values $d_{0}=1/12,$ $%
d_{2}=d_{3}=1,$ $q=-1,$ $Q=\alpha ,$ $R_{0}=1$ and $V_{0}=\alpha Ze^{2}$ in
Eq. (29)$,$ then we obtain the energy spectrum of the Hulth\'{e}n potential:%
\begin{equation}
E_{n,l}^{(D)}=\frac{\hbar ^{2}\alpha ^{2}}{2\mu }\left\{ \frac{1}{12}\left(
l+\frac{D-1}{2}\right) \left( l+\frac{D-3}{2}\right) -\left[ \frac{\mu Ze^{2}%
}{\hbar ^{2}\left( n+l+\frac{D-1}{2}\right) \alpha }-\frac{\left( n+l+\frac{%
D-1}{2}\right) }{2}\right] ^{2}\right\} .
\end{equation}%
which is similar to the one obtained in Ref. [1] (if $d_{0}=0$ in the usual
approximation).

\section{Eigenfunctions}

We are now in the position to study the corresponding eigenfunction of this
quantum system for completeness. The Riccati equation of the relation (8) is%
\begin{equation}
\phi {}^{\prime }(r)=-\frac{2\mu }{\hbar ^{2}}\left[ E_{nl}-V_{eff}(r)\right]
-\phi (r)^{2},
\end{equation}%
where%
\begin{equation}
\phi (r)=\frac{u{}^{\prime }(r)}{u(r)}.
\end{equation}%
Based on
\begin{equation}
u(r)=e^{\int^{r}\phi (r)dr}=e^{-a\int^{r}\frac{1}{y\left( 1-qy\right) }\phi
(z)dz},
\end{equation}%
and using Eq. (24), we can easily calculate the eigenfunction of the ground
state as%
\begin{equation}
u_{0}(r)=N_{0}\left( e^{-\frac{\left( r-R_{0}\right) }{a}}\right) ^{%
\widetilde{\varepsilon }_{0}}\left( 1+qe^{-\frac{\left( r-R_{0}\right) }{a}%
}\right) ^{\nu },\text{ }\widetilde{\varepsilon }_{0}>0,\text{ }\nu \geq 1,
\end{equation}%
where
\begin{equation}
\widetilde{\varepsilon }_{n=0}=a\sqrt{\frac{2\mu }{\hbar ^{2}}\left(
bd_{0}-E_{0}\right) },\text{ }\nu =\frac{1}{2}\left( 1-\sqrt{1+\frac{8\mu }{%
\hbar ^{2}}\frac{Q^{2}\delta ^{2}}{q^{2}\alpha ^{2}}d_{2}}\right) ,
\end{equation}%
and $N_{0}$ is the normalization constant.

Let us find the eigenfunction for any quantum state $n$ other than ground
state$.$ By considering the boundary conditions%
\begin{equation}
y=\left\{
\begin{array}{ccc}
0 & \text{when} & r\rightarrow \infty , \\
1 & \text{when} & r\rightarrow 0,%
\end{array}%
\right.
\end{equation}%
with $u(y)\rightarrow 0,$ we may define a more general radial
eigenfunctions, valid for any quantum number $n,$ of the form:%
\begin{equation}
u(y)=y^{\widetilde{\varepsilon }_{n,l}}\left( 1+qy\right) ^{\nu }f(y),\text{
}y=e^{-\frac{\left( r-R_{0}\right) }{a}},\text{ }\widetilde{\varepsilon }%
_{n,l}>0,\text{ }\nu \geq 1.
\end{equation}%
Substituting Eq. (36) into Eq. (8) leads to the following hypergeometric
equation%
\begin{equation*}
y\left( 1+qy\right) f^{\prime \prime }(y)+\left[ 1+2\widetilde{\varepsilon }%
_{n,l}+q\left( 1+2\widetilde{\varepsilon }_{n,l}+2\nu \right) y\right]
f{}^{\prime }(y)
\end{equation*}%
\begin{equation}
-\left[ \frac{2\mu }{\hbar ^{2}}a^{2}\left( bd_{1}-V_{0}\right) -q\nu \left(
1+2\widetilde{\varepsilon }_{n,l}\right) \right] f(y)=0,
\end{equation}%
with the following conditions:
\begin{subequations}
\begin{equation}
\widetilde{\varepsilon }_{n,l}^{2}+\frac{2\mu }{\hbar ^{2}}a^{2}\left(
E_{n,l}-bd_{0}\right) =0,
\end{equation}%
\begin{equation}
q^{2}\nu (\nu -1)-\frac{2\mu }{\hbar ^{2}}a^{2}bd_{2}=0,
\end{equation}%
which leads to the following results
\end{subequations}
\begin{subequations}
\begin{equation}
\widetilde{\varepsilon }_{n,l}=a\sqrt{\left( l+\frac{D-1}{2}\right) \left( l+%
\frac{D-3}{2}\right) \frac{Q^{2}d_{0}}{R_{0}^{2}}-\frac{2\mu }{\hbar ^{2}}%
a^{2}E_{n,l}},
\end{equation}%
\begin{equation}
\nu =\frac{1}{2}\left( 1\pm \zeta \right) ,\text{ }\zeta =\sqrt{1+4\left( l+%
\frac{D-1}{2}\right) \left( l+\frac{D-3}{2}\right) \frac{Q^{2}}{q^{2}\alpha
^{2}}d_{2}},
\end{equation}%
where $+$ and $-$ are taken for the Hulth\'{e}n and the Woods-Saxon
potentials, respectively. The solution of the Eq. (40) is then
\end{subequations}
\begin{equation}
F(y)=%
\begin{array}{c}
_{2}F_{1}%
\end{array}%
\left( A,B;C;y\right) =\frac{\Gamma (C)}{\Gamma (A)\Gamma (B)}%
\sum\limits_{k=0}^{\infty }\frac{\Gamma (A+k)\Gamma (B+k)}{\Gamma (C+k)}%
\frac{y^{k}}{k!},
\end{equation}%
where%
\begin{equation*}
A=\widetilde{\varepsilon }_{n,l}-q\nu -\sqrt{U}=-n,\text{ }n=0,1,2,\cdots ,
\end{equation*}%
\begin{equation}
B=\widetilde{\varepsilon }_{n,l}-q\nu +\sqrt{U},
\end{equation}%
\begin{equation*}
C=1+2\widetilde{\varepsilon }_{n,l},
\end{equation*}%
where $%
\begin{array}{c}
_{2}F_{1}%
\end{array}%
\left( A,B;C;y\right) $\ is the hypergeometric function. Now, we may write
down the radial wave functions (34) as%
\begin{equation}
u_{n,l}(r)=\mathcal{N}_{nl}\left( e^{-\frac{\left( r-R_{0}\right) }{a}%
}\right) ^{\widetilde{\varepsilon }_{n,l}}\left( 1+qe^{-\frac{\left(
r-R_{0}\right) }{a}}\right) ^{\nu }%
\begin{array}{c}
_{2}F_{1}%
\end{array}%
\left( -n,n+2\widetilde{\varepsilon }_{n,l}+q\zeta -q;1+2\widetilde{%
\varepsilon }_{n,l};e^{-\frac{\left( r-R_{0}\right) }{a}}\right) .
\end{equation}%
If now, we set $n=0$ in Eq. (45)$,$ then we can easily obtain Eq. (36).
Finally, the unnormalized total wave functions are obtained as%
\begin{equation*}
\psi _{n,l,m}(r,\Omega _{D})=\mathcal{N}_{nl}r^{-(D-1)/2}\left( e^{-\frac{%
\left( r-R_{0}\right) }{a}}\right) ^{\widetilde{\varepsilon }_{n,l}}\left(
1+qe^{-\frac{\left( r-R_{0}\right) }{a}}\right) ^{\nu }
\end{equation*}%
\begin{equation}
\times
\begin{array}{c}
_{2}F_{1}%
\end{array}%
\left( -n,n+2\widetilde{\varepsilon }_{n,l}+q\zeta -q;1+2\widetilde{%
\varepsilon }_{n,l};e^{-\frac{\left( r-R_{0}\right) }{a}}\right)
Y_{l}^{m}(\Omega _{D}).
\end{equation}%
which is identical to Eq. (42) of Ref. [33] when $D=3.$ Thus, the Jacobi
polynomials can be expressed in terms of the hypergeometric functions [34]
\begin{equation}
P_{n}^{\left( A,B\right) }(1-2x)=\frac{\Gamma \left( n+1+A\right) }{n!\Gamma
\left( 1+A\right) }%
\begin{array}{c}
_{2}F_{1}%
\end{array}%
(-n,n+A+B+1;A+1;x).
\end{equation}%
It is worth noting that the hypergeometric function $_{2}F_{1}(A,B;C;x)$ is
a special case of the generalized hypergeometric function [34]%
\begin{equation}
_{p}F_{q}(\alpha _{1},\alpha _{2},\cdots ,\alpha _{p};\beta
_{1},\beta _{1},\cdots ,\beta _{q};x)=\sum\limits_{k=0}^{\infty
}\frac{\left( \alpha _{1}\right) _{k}\left( \alpha _{2}\right)
_{k}\cdots \left( \alpha _{p}\right) }{\left( \beta _{1}\right)
_{k}\left( \beta _{2}\right) _{k}\cdots \left( \beta _{q}\right)
}\frac{x^{k}}{k!},
\end{equation}%
where the Pochhammer symbol is defined by $\left( y\right) _{k}=\Gamma
(y+k)/\Gamma (y).$

\section{Interdimensional Degeneracy}

From Eq. (30), it can be seen that two interdimensional states are
degenerate whenever [35]%
\begin{equation}
(n,l,D)\rightarrow (n,l\pm 1,D\mp 2)\text{ }\Rightarrow \text{ }%
E_{n,l}^{(D)}\rightarrow E_{n,l\pm 1}^{(D\mp 2)},
\end{equation}%
i.e,%
\begin{equation}
\text{ }E_{n,l}^{(D)}=E_{n,l\pm 1}^{(D\mp 2)}=\frac{\delta ^{2}}{R_{0}^{2}}%
d_{0}-\frac{\hbar ^{2}a^{2}}{2\mu }\left[ \frac{\left( 2n+1-\sqrt{1+\frac{%
8\mu }{\hbar ^{2}}\frac{\delta ^{2}}{\alpha ^{2}}d_{2}}\right) }{4a^{2}}-%
\frac{\frac{2\mu }{\hbar ^{2}}\frac{\delta ^{2}}{R_{0}^{2}}\left(
d_{1}+d_{2}\right) -\frac{2\mu }{\hbar ^{2}}V_{0}}{\left( 2n+1-\sqrt{1+\frac{%
8\mu }{\hbar ^{2}}\frac{\delta ^{2}}{\alpha ^{2}}d_{2}}\right) }\right] ^{2}.
\end{equation}%
Thus, a knowledge of $E_{n,l}^{(D)}$ for $D=2$ and $D=3$ provides the
information necessary to find $E_{n,l}^{(D)}$ for other higher dimensions.

For example, $E_{0,4}^{(2)}=E_{0,3}^{(4)}=E_{0,2}^{(6)}=E_{0,1}^{(8)}.$ This
is the same transformational invariance described for bound states of free
atoms and molecules [36-38] and demonstrates the existence of
interdimensional degeneracies among states of the confined Hulth\'{e}n
potential. \

\section{Conclusions}

By using the improved quantization rule, we have given an alternative method
to obtain the approximated energy eigenvalues and eigenfunctions of the Schr%
\"{o}dinger equation in $D$-dimensions with the Woods-Saxon potential for
all $l$-states within the Pekeris approximation to deal with centrifugal
term. We emphasize that the expressions obtained for the energy eigenvalues
and eigenfunctions are valid for all real values of $R_{0}.$ However, the
solutions provided in Refs. [10,11] are at most valid for $R_{0}=0$
(Rosen-Morse-type potential and not the Woods-Saxon potential). The
advantage of this method is that it gives the eigenvalues through the
calculation of two integrations (26) and (27) and solving the resulting
algebraic equation. First, we can easily obtain the quantum correction by
only considering the solution of the ground state of quantum system since it
is independent of the number of nodes of wave function for exactly solvable
quantum system. Second, the wave functions have also been obtained by
solving the Riccati equation. The general expressions obtained for the
energy eigenvalues and wave functions can be easily reduced to the
Woodes-Saxon and the Hulthen potentials, the three-dimensional space ($D=3$%
), $s$-wave ($l=0$) cases. The method presented here is a systematic
one, simple, practical and powerful than the other known methods.
Finally, the simplicity of the method motivates us it to solve the
Dirac equation for this WS potential to see the resulting
relativistic effects.

\acknowledgments The partial support provided by the Scientific and
Technological Research Council of Turkey is highly appreciated.

\appendix

\section{Integral Formulas}

The following integral formulas are useful during the calculation of the
momentum integral and the quantum correction term:\

\begin{equation}
\int\limits_{r_{A}}^{r_{B}}\frac{r}{\sqrt{(r-r_{A})(r_{B}-r)}}dr=\frac{\pi }{%
2}(r_{A}+r_{B}),
\end{equation}

\begin{equation}
\int\limits_{r_{A}}^{r_{B}}\frac{1}{r\sqrt{(r-r_{A})(r_{B}-r)}}dr=\frac{\pi
}{\sqrt{r_{A}r_{B}}},
\end{equation}%
\begin{equation}
\int\limits_{r_{A}}^{r_{B}}\frac{1}{\sqrt{(r-r_{A})(r_{B}-r)}}dr=\pi ,
\end{equation}%
\begin{equation}
\int\limits_{r_{A}}^{r_{B}}\frac{1}{r}\sqrt{(r-r_{A})(r_{B}-r)}dr=\pi \left[
\frac{1}{2}(r_{A}+r_{B})-\sqrt{r_{A}r_{B}}\right] ,.
\end{equation}%
\begin{equation}
\int\limits_{r_{A}}^{r_{B}}\frac{1}{(a+br)\sqrt{(r-r_{A})(r_{B}-r)}}dr=\frac{%
\pi }{\sqrt{(a+br_{A})(a+br_{B})}},\text{ }r_{B}>r_{A}>0.
\end{equation}

\newpage

{\normalsize 
}

\bigskip

\bigskip

\bigskip


\begin{thebibliography}{99}
\bibitem{1} X. -Y. Gu and J. -Q. Sun, J. Math. Phys. 51 (2), 022106 (2010).

\bibitem{2} R. D. Woods and D. S. Saxon, Phys. Rev. 95, 577 (1954).

\bibitem{3} A. K. Bose, Nuovo Cimento 32, 679 (1964).

\bibitem{4} S. Fl\"{u}gge, Practical Quantum Mechanics, Springer-Verlag,
Berlin, 1974.

\bibitem{5} J. M. G. G\^{o}mez, K. Kar, V. K. B. Kota, R. A. Molina and J.
Retamosa, Phys. Lett. B 567, 251 (2003).

\bibitem{6} S. E. Massen and C. P. Panos, Phys. Lett. A 246, 530 (1997).

\bibitem{7} B. A. Kotsos and M. E. Grypeos, Physica B 229, 173 (1997).

\bibitem{8} H. Nicolai, J. Phys. A 9, 1497 (1976).

\bibitem{9} O. V. Bespulova, E. A. Romanovsky and T. I. Spasskaya, J. Phys.
G 29, 1193 (2003).

\bibitem{10} C. Berkdemir, A. Berkdemir and R. Sever, Phys. Rev. C 72,
027001 (2005).

\bibitem{11} C. Berkdemir, A. Berkdemir and R. Sever, Phys. Rev. C 74,
039902(E) (2006).

\bibitem{12} S. M. Ikhdair and R. Sever, Int. J. Theor. Phys. 46, 1643
(2007).

\bibitem{13} C. Berkdemir, A. Berkdemir and R. Sever, J. Math. Chem. 43, 944
(2008).

\bibitem{14} P. Kennedy, J. Phys. A 35, 689 (2002).

\bibitem{15} C. Berkdemir, A. Berkdemir and R. Sever, J. Phys. A 39, 13455
(2006).

\bibitem{16} J. -L. Tian, N. Wang and Z. -X. Li, Chinese Phys. Lett. A 24,
905 (2007).

\bibitem{17} J. -Y. Guo and Z. -Q. Sheng, Phys. Lett. A 338, 90 (2005).

\bibitem{18} C. Rojas and V. M. Villalba, Phys. Rev. A 71, 052101 (2005).

\bibitem{19} O. Aydo\u{g}du and R. Sever, Eur. Phys. J. A 43, 73 (2010).

\bibitem{20} Z. Q. Ma and B. W. Xu, Europhys Lett. 69, 685 (2005).

\bibitem{21} Z. Q. Ma and B. W. Xu, Int. J. Mod. Phys. E 14, 599 (2005).

\bibitem{22} W. -C. Qiang and S. -H. Dong, Phys. Lett. A 363, 169 (2007).

\bibitem{23} W. -C. Qiang, R. -S. Zhou and Y. Gao, J. Phys. A: Math. Theor.
40, 1677 (2007).

\bibitem{24} S. M. Ikhdair and R. Sever, J. Math. Chem. 45 (4), 1137 (2009).

\bibitem{25} Z. -Q. Ma, A. Gonzalez-Cisneros, B. -W. Xu and S. -H. Dong,
Phys. Lett. A 371, 180 (2007).

\bibitem{26} S. -H. Dong and A. Gonzalez-Cisneros, Ann. Phys. 323, 1136
(2008).

\bibitem{27} R. L. Greene and C. Aldrich, Phys. Rev. A 14, 2363 (1976).

\bibitem{28} S. M. Ikhdair and R. Sever, J. Math. Chem. 42, 461 (2007).

\bibitem{29} N. Saad, Phys. Scr. 76, 623 (2007).

\bibitem{30} C. L. Pekeris, Phys. Rev. 45, 98 (1934).

\bibitem{31} C. Berkdemir and J. Han, Chem. Phys. Lett. 409, 203 (2005).

\bibitem{32} C. Berkdemir, Nucl. Phys. A 770, 32 (2006).

\bibitem{33} S.M. Ikhdair and R. Sever, Cent. Eur. J. Phys. 8 (4), 652
(2010).

\bibitem{34} I.S. Gradshteyn and I.M Ryzhik, Tables of Integrals, Series,
and Products, 5th edn (New York, Academic, 1994).

\bibitem{35} H. E. Montgomery Jr., N. A. Aquino and K. D. Sen, Int. J.
Quantum Chem. 107, 798 (2007).

\bibitem{36} D. R. Herrick, J. Math. Phys. 16, 281 (1975).

\bibitem{37} D. R. Herrick and F. H. Stillinger, Phys. Rev. 11, 42 (1975).

\bibitem{38} D. D. Fratz and D. R. Herschbach, \ J. Chem. Phys. 92, 6668
(1990).
\end{thebibliography}
\end{document}